\documentclass[journal,transmag]{IEEEtran}
\usepackage{cite,bm,amssymb,amsthm,url,multirow,times,enumitem,comment}

\usepackage{subfigure,wrapfig,graphicx,booktabs,fancyhdr,amsmath,amsfonts}

\newcommand{\ba}{\begin{array}}
\newcommand{\ea}{\end{array}}

\DeclareMathAlphabet{\mathpzc}{OT1}{pzc}{m}{it}

\ifCLASSINFOpdf
\else
\fi
\begin{document}
%
\title{Deep urban unaided precise \\ Global Navigation Satellite System vehicle positioning}


\author{\IEEEauthorblockN{Todd E. Humphreys\IEEEauthorrefmark{1},~\IEEEmembership{Member,~IEEE,}
Matthew J. Murrian\IEEEauthorrefmark{1},~\IEEEmembership{Member,~IEEE,} and \\
Lakshay Narula\IEEEauthorrefmark{2},~\IEEEmembership{Student Member,~IEEE}}
\IEEEauthorblockA{\IEEEauthorrefmark{1}Aerospace Engineering and Engineering Mechanics,
The University of Texas at Austin, Austin, TX 78712 USA.}
\IEEEauthorblockA{\IEEEauthorrefmark{2}Electrical and Computer Engineering,
The University of Texas at Austin, Austin, TX 78712 USA.}
}

\IEEEtitleabstractindextext{%
  \begin{abstract}
    This paper presents the most thorough study to date of vehicular
    carrier-phase differential GNSS (CDGNSS) positioning performance in a deep
    urban setting unaided by complementary sensors.  Using data captured
    during approximately 2 hours of driving in and around the dense urban
    center of Austin, TX, a CDGNSS system is demonstrated to achieve
    17-cm-accurate 3D urban positioning (95\% probability) with solution
    availability greater than 87\%.  The results are achieved without any
    aiding by inertial, electro-optical, or odometry sensors.  Development and
    evaluation of the unaided GNSS-based precise positioning system is a key
    milestone toward the overall goal of combining precise GNSS, vision,
    radar, and inertial sensing for all-weather high-integrity
    high-absolute-accuracy positioning for automated and connected vehicles.
    The system described and evaluated herein is composed of a densely-spaced
    reference network, a software-defined GNSS receiver, and a real-time
    kinematic (RTK) positioning engine.  A performance sensitivity analysis
    reveals that navigation data wipeoff for fully-modulated GNSS signals
    and a dense reference network are key to high-performance urban RTK
    positioning.  A comparison with existing unaided systems for urban GNSS
    processing indicates that the proposed system has significantly greater
    availability or accuracy.
\end{abstract}

}

\maketitle

\IEEEdisplaynontitleabstractindextext

%
\IEEEpeerreviewmaketitle

\newif\ifpreprint
\preprintfalse
\preprinttrue

\ifpreprint

\pagestyle{plain}
\thispagestyle{fancy}  
\fancyhf{} 
\renewcommand{\headrulewidth}{0pt}
\rfoot{\footnotesize \bf Preprint of the 2020 IEEE ITS Magazine} \lfoot{\footnotesize \bf
  Copyright \copyright~2020 by Todd Humphreys, Matthew Murrian, \\ and Lakshay Narula}
\else

\thispagestyle{empty}
\pagestyle{empty}

\fi


\section{Introduction}
%
%
%
%
\IEEEPARstart{F}{uture} Vehicle-to-vehicle (V2V) and vehicle-to-infrastructure
(V2I) connectivity will permit vehicles to relay their positions and
velocities to each other with millisecond latency, enabling tight coordinated
platooning and efficient intersection management.  More ambitiously, broadband
V2V and V2I enabled by 5G wireless networks will permit vehicles to share
unprocessed or lightly-processed sensor data. \emph{Ad hoc} networks of
vehicles and infrastructure will then function as a single sensing organism.
The risk of collisions, especially with pedestrians and cyclists---notoriously
unpredictable and much harder to sense reliably than vehicles---will be
significantly reduced as vehicles and infrastructure contribute sensor data
from multiple vantage points to build a blind-spot-free model of their
surroundings.

Such collaborative sensing and traffic coordination requires vehicles to know
and share their own position.  How accurately?  The proposed Dedicated Short
Range Communications (DSRC) basic safety message, a first step in V2V
coordination, does not yet define a position accuracy requirement, effectively
accepting whatever accuracy a standard GNSS receiver provides
\cite{kenney2011dedicated}.  But automated intersection management
\cite{fajardo2011automated}, tight-formation platooning, and unified
processing of sensor data---all involving vehicles of different makes that may
not share a common map---will be greatly facilitated by globally-referenced
positioning with sub-30-cm accuracy.

Poor weather also motivates high-accuracy absolute positioning.  Every
automated vehicle initiative of which the present authors are aware depends
crucially on lidar or cameras for fine-grained positioning within their local
environment.  But these sensing modalities perform poorly in low-visibility
conditions such as a snowy whiteout, dense fog, or heavy rain.  Moreover,
high-definition 3D maps created with lidar and camera data, maps that have
proven crucial to recent progress in reliable vehicle automation, can be
rendered dangerously obsolete by a single snowstorm, leaving vehicles who rely
on such maps for positioning no option but to fall back on GNSS and radar to
navigate a snow-covered roadway in low-visibility conditions.  When, as is
often the case on rural roads, such snowy surroundings offer few
radar-reflective landmarks, radar too becomes useless.  GNSS receivers operate
well in all weather conditions, but only a highly accurate GNSS solution,
e.g., one whose absolute errors remain under 30 cm 95\% of the time, could
prevent a vehicle's drifting onto a snow-covered road's soft shoulder.  Code-
and Doppler-based GNSS solutions can be asymptotically accurate (averaged over
many sessions) to better than 50 cm, which may be adequate for digital mapping
\cite{narula2018accurate}, but they will find it challenging to meet a 30 cm
95\% stand-alone requirement, even with modernized GNSS offering wideband
signals at multiple frequencies.

Carrier-phase-based GNSS positioning---also referred to as precise GNSS
positioning even though it actually offers absolute accuracy, not just
precision (repeatability)---can meet the most demanding accuracy requirements
envisioned for automated and connected vehicles, but has historically been
either too expensive or too fragile, except in open areas with a clear view of
the overhead satellites, for widespread adoption.  Coupling a carrier-phase
differential GNSS (CDGNSS) receiver with a tactical grade inertial sensor, as
in \cite{petovello2004benefits,scherzinger2006precise,
  zhangComparisonWithTactical2006,kennedy2006architecture} enables robust
high-accuracy positioning even during the extended signal outages common in
dense urban areas.  But GNSS-inertial systems with tactical-grade inertial
measurement units (IMUs) cost tens of thousands of dollars and have proven
stubbornly resistant to commoditization.  Coupling a GNSS receiver with
automotive- or industrial-grade IMUs is much more economical, and
significantly improves performance, as shown in \cite{li2018high}.  But such
coupling only allows approximately 5 seconds of complete GNSS signal blockage
before the IMU no longer offers a useful constraint for so-called integer
ambiguity resolution \cite{evaluationLowCostMems2006Godha}, which underpins
the fastest, most accurate, and most robust CDGNSS techniques, namely,
single-baseline RTK, network RTK, and PPP-RTK
\cite{teunissen2015review,cui2017rtkforCAV}.

Previous research has suggested an inexpensive technique for robustifying RTK
positioning: tightly coupling carrier-phase-based GNSS positioning with
inertial sensing and vision \cite{shepard2014fusion,pesyna2015dissertation}.
Such coupling takes advantage of the remarkable progress in high-resolution,
low-cost cameras within the intensely competitive smartphone market.  The
current authors are engaged in developing a high-integrity RTK-vision system
for high-accuracy vehicular positioning in rural and urban environments.
Further coupling with radar will make the system robust to low-visibility
conditions.

As a step toward this goal, it is of interest to evaluate the performance of
stand-alone RTK techniques---those unaided by IMUs, odometry, or vision---in
urban environments.  Such a study will reveal why and when aiding is
necessary, and how an RTK positioning system might behave if aiding were
somehow impaired or unavailable, whether due to sensor faults or, in the case
of exclusive visual aiding, poor visibility conditions.

Little prior work has explored unaided vehicular RTK performance in urban
environments, no doubt because performance results have historically been
dismal. Short-baseline RTK experiments between two vehicles in
\cite{ong2009assessment} revealed that multi-frequency (L1-L2) GPS and Glonass
RTK yielded poor results in residential and urban environments.  Only along a
mountain highway with a relatively clear view of the sky was availability
greater than 90\% and accuracy better than 30 cm.  RTK positioning in downtown
Calgary was disastrous, with less than 60\% solution availability and RMS
errors exceeding 9 meters.

More recently, Li et al. \cite{li2018high} have shown that, with the benefit
of greater signal availability, unaided professional-grade dual-frequency GPS
+ BDS + GLONASS RTK can achieve correct integer fixing rates of 76.7\% on a
1-hour drive along an urban route in Wuhan, China.  But Li et al. do not
provide data on the incorrect fixing rate, nor a full error distribution, so
the significance of their results is difficult to assess.

Recent urban RTK testing by Jackson et al. \cite{jackson2018assessmentRtk}
indicates that no low-to-mid-range consumer RTK solution offers greater than
35\% fixed (integer-resolved) solution availability in urban areas, despite a
dense reference network and dual-frequency capability.  A key failing of
existing receivers appears to be their slow recovery after passing under
bridges or overpasses.

This paper describes and evaluates an unaided RTK positioning system that has
been designed for vehicular operation in both rural and urban environments.
Preliminary performance results were published in a conference version of this
paper \cite{humphreys2018urbanStandAloneRTKplans2018}.  The current paper
improves on the conference version in four ways: (1) the test route is both
more challenging and more comprehensive, (2) a proper independent ground truth
trajectory is used as the basis of error evaluation, (3) data modulation
wipeoff for improved carrier tracking robustness is applied not only on GPS L1
C/A signals, as previously, but now also on SBAS L1 signals, and (4) the
performance benefit of vehicle GNSS antenna calibration is assessed.

This paper's primary contributions are (i) a demonstration of the performance
that can be achieved with a low-cost software-defined unaided RTK GNSS
platform in a dense urban environment, and (ii) an evaluation of the relative
importance of various factors (e.g., data bit wipeoff, age of reference data,
rover antenna calibration, reference network density) to the overall system
performance.

To stimulate further innovation in urban precise positioning, all data from
this paper's urban driving campaign have been posted at
\url{http://radionavlab.ae.utexas.edu} under ``Public Datasets,'' including
wideband (10 MHz) intermediate frequency samples from both the reference and
rover antennas, RINEX-formatted rover and reference observables, and the
ground truth trajectory.

\section{Challenges of mobile precise positioning in Urban Environments}
The mobile urban satellite-to-user channel is distinguished by rapid channel
evolution.  As the vehicle travels along streets closely lined with tall
buildings, only glimpses of power are available from signals arriving from
directions roughly perpendicular to the roadway.  A GNSS receiver designed to
provide phase-locked carrier measurements for RTK positioning in such
environments must simultaneously (1) prevent frequency unlock during the deep
fades caused by building occlusions, and (2) exploit momentary signal
availability by immediately acquiring full-cycle phase lock and indicating
this to downstream processing.

Tracking in the mobile urban channel is unlike indoor or weak-signal tracking,
such as explored in \cite{m_psiaki02_wgs,niedermeier2010dingpos}, in that the
urban fading environment is substantially binary: either the line-of-sight
signal is present at a fairly healthy carrier-to-noise ratio $C/N_0$, or it is
hopelessly attenuated after passing through entire buildings constructed of
concrete, steel, and glass.  The traditional weak-signal-tracking technique of
extending the signal integration time and lowering the tracking loop
bandwidths can be useful to slow the rate of frequency unlock during such
fading, but not for actually recovering a weak signal from the noise.  There
is simply no signal to recover.

\begin{figure}[!t]
\centering
\includegraphics[width = 8.5cm]{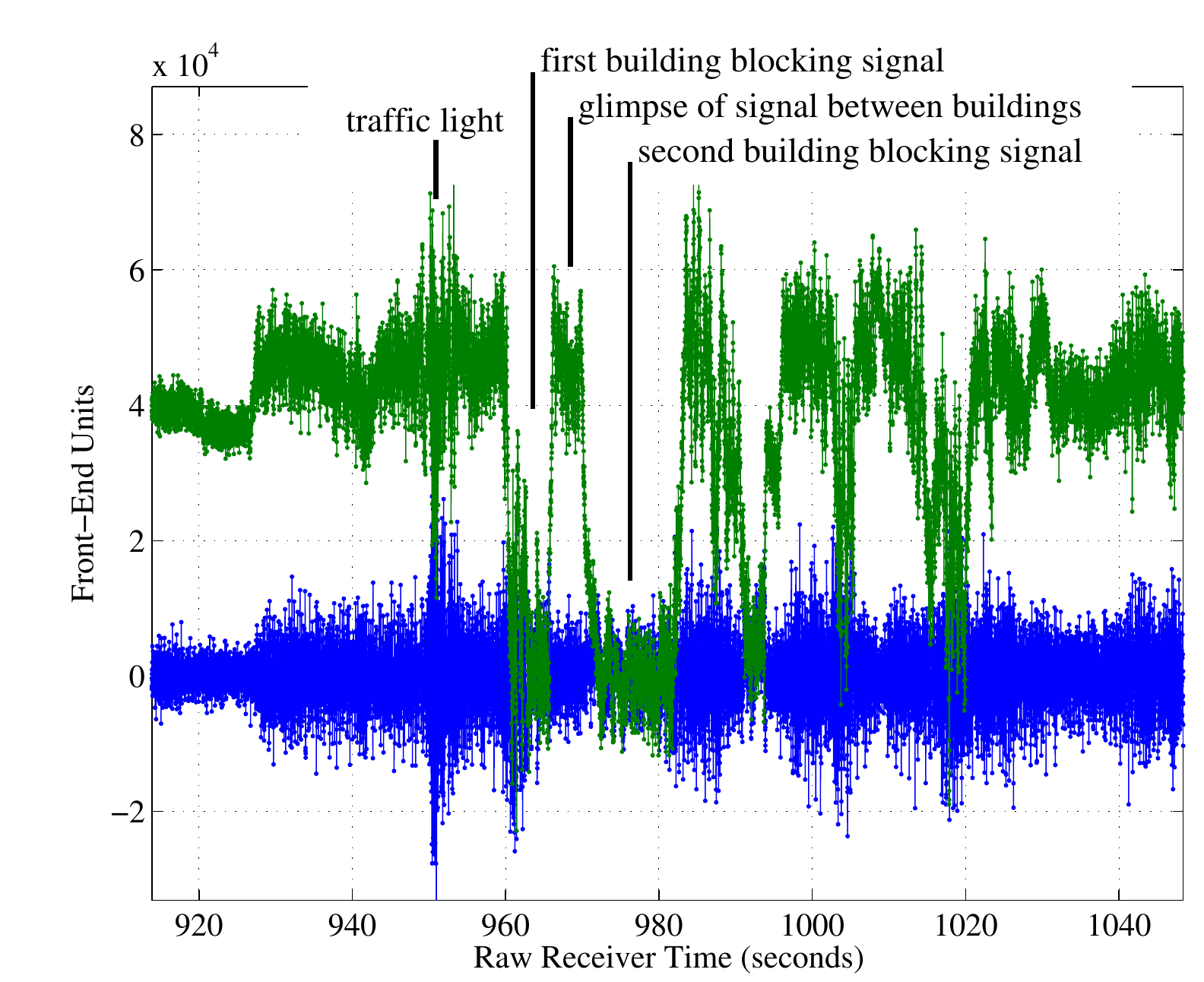}
\caption{In-phase (green, top) and quadrature (blue, bottom) 10-ms complex
  correlation products for a GPS L1 C/A signal at 35 degrees elevation
  arriving from the south to a vehicle traveling west on an urban roadway.
  The 20-ms LNAV navigation data bits have been wiped off to allow full
  carrier cycle recovery.  Rapid fading---and rapid recovery---occur as
  buildings intermittently block the signal.}
\label{fig:iq_gps_prn4P}
\end{figure}

Fig. \ref{fig:iq_gps_prn4P} illustrates this fact.  The initial disturbance at
950 seconds is due to an overhead traffic light.  This is followed in rapid
succession by a complete signal blockage due to a tall building on the south
side of the east-west street, a brief (four-second) interval of clear
satellite availability as the receiver catches a glimpse of the signal between
two buildings, and another signal eclipse by a second building.

A GNSS receiver designed for urban tracking will make full use of such
between-building glimpses.  This requires immediate (within approximately 100
ms) recovery of full-cycle phase lock, which is only possible on
suppressed-carrier signals like GPS L1 C/A if the receiver can accurately
predict the modulating data symbols.  Downstream RTK processing must also be
poised to exploit signal glimpses by identifying and rejecting observables
from blocked or otherwise compromised signals, and by immediately
re-evaluating the corresponding integer ambiguities when signals reappear.  A
multi-stage cycle slip detection and recovery technique, such as proposed in
\cite{s_mohiuddin07_wia}, is too slow for urban positioning.

A related hallmark of the urban mobile channel is the wide and rapid variation
of the number of signals available for RTK positioning.  The number
$N_{\rm DD}$ of double-difference (DD) signals (each one providing a DD
pseudorange and a DD carrier phase observable) varies widely whenever the
vehicle is moving.  The implication for RTK processing is that integer
ambiguity continuity will often be lost, requiring rapid and continuous
re-estimation of ambiguities.

\section{System Description}

\subsection{Overview}
\begin{figure}[!t]
\centering
\includegraphics[width = 8.5cm]{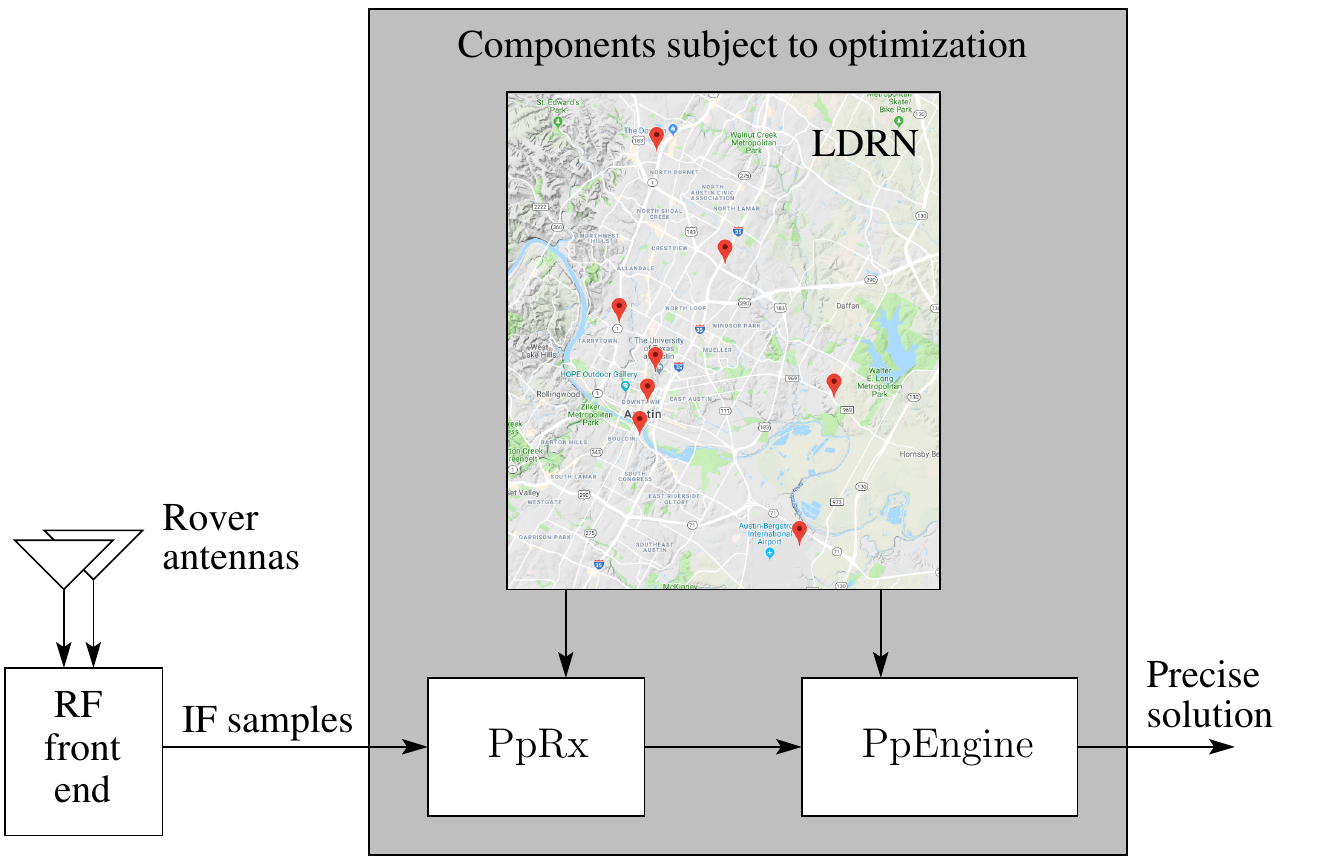}
\caption{The University of Texas precise positioning system.}
\label{fig:pipeline}
\end{figure}

GNSS components of this paper's precise positioning system are shown in
Fig. \ref{fig:pipeline}.  The sub-components enclosed in the gray box are the
target of the present work's optimization efforts for good performance in
urban environments.

Two rover antennas feed analog signals to a radio frequency (RF) front end,
which down-mixes and digitizes the signals, producing a stream of intermediate
frequency (IF) samples.  The RF front end used in the present work produces
samples at 10 MHz for two antennas and two frequencies: a band centered at GPS
L1 and one centered at GPS L2.  The (single-sided) analog bandwidth of each
band is 4 MHz---wide enough to capture over 90\% of the power in the GPS L1
C/A, Galileo E1 BOC(1,1), and GPS L2C signals.

Four IF sample streams, one for each antenna and band, are fed to PpRx, an
embeddable multi-frequency software-defined GNSS receiver developed primarily
at The University of Texas
\cite{t_humphreys06_scp,t_humphreys09_sdgr,lightsey2013demonstration}.  PpRx
draws ephemeris data, GPS LNAV and SBAS (WAAS) data bit estimates from the
Longhorn Dense Reference Network (LDRN), a set of 8 GNSS reference stations
deployed in Austin, TX.  Each reference station in the LDRN runs a
strict-real-time variant of PpRx and sends its data to a central network
server from which any compatible receiver can draw assistance data and network
observables. 

PpRx feeds code and carrier observables, and other useful signal information,
to an RTK engine called PpEngine.  For the results presented in this paper,
PpEngine draws observables and ephemeris data from a single LDRN reference
station at a time---the traditional RTK topology.  The precise solution
produced by PpEngine is a fixed (integer-resolved) or float solution depending
on the results of an integer aperture test \cite{teunissen2003integer}.

\subsection{Performance Metrics}
\label{sec:performance-metrics}
The performance of precise positioning systems in safety-of-life applications
is assessed in terms of integrity, accuracy, and availability
\cite{green2018iagiab,green2018pdigiab}.  For several emerging applications of
practical interest, such as automated and connected vehicles, no regulatory
body has set clear positioning performance requirements.  An industry
consensus appears to be emerging which calls for a 95\% accuracy requirement
of 30 cm, but it is not clear what the associated integrity risk or continuity
requirements should be.  It is likely that the U.S. National Highway Traffic
Safety Administration, and other regulatory bodies worldwide, will eventually
issue positioning performance requirements for connected and automated
vehicles.

This paper focuses on four related performance metrics: $d_{95}$, the 95th
percentile error magnitude for fixed solutions, $P_V$, the probability (or
rate in continuous trials) that a validated (fixed) solution is available at
each epoch, as opposed to a fallback float solution, $P_S$, the probability of
correctly (successfully) resolving the full integer set at each epoch, and
$P_F$, the probability that one or more integer ambiguities failed to resolve
correctly at each epoch \cite{green2018iagiab}.  $P_V$, $P_S$, and $P_F$ are
related by $P_V = P_S + P_F$.  A fourth probability,
$P_U = 1 - P_F - P_S = 1-P_V$, that of the undecided event, is the probability
that a float solution, or no solution at all, is produced, due to an aperture
test failure or failure of some other validation test.

An unavoidable tradeoff between $P_S$ and $P_F$ exists such that any widening
of the integer aperture region to increase $P_S$ comes at the expense of an
increase in $P_F$ (not necessarily of the same amount)
\cite{teunissen2005iab}.  Therefore, an optimization problem can be stated in
terms of $P_S$ and $P_F$ as follows: maximize $P_S$ for $P_F \leq \bar{P}_F$,
where $\bar{P}_F$ is a fixed tolerable probability of failed fixing.  Integer
aperture bootstrapping techniques such as \cite{teunissen2005iab} and its
generalization to partial ambiguity resolution in \cite{green2018iagiab}
analytically determine thresholds for the integer aperture test to ensure
$P_F \leq \bar{P}_F$.  For the optimal integer least squares (ILS) approach
adopted in this paper, it is not possible to calculate an analytical aperture
threshold, but an approximate one can be obtained via simulation such that
$P_F \leq \bar{P}_F$ is satisfied almost surely \cite{wang2015iafit}.  A value
of $\bar{P}_F = 0.001$ was adopted for the present paper, meaning that a
fixing failure rate less than 1 in 1000 epochs was deemed acceptable.
However, multipath, GNSS signal passage through foliage, and other signal
impairments common in urban areas cause the empirical $P_F$ to significantly
exceed $\bar{P}_F$ when the aperture threshold is chosen according to the
Gaussian error assumptions ubiquitous in the integer aperture
literature. Thus, a looser \emph{empirical} upper bound $\bar{\bar{P}}_F$ must
be chosen.  The optimization problem is then to maximize $P_S$ subject to the
empirical $P_F$ respecting the bound $P_F \leq \bar{\bar{P}}_F$.

\subsection{Design Philosophy}
With origins in scintillation-resistant carrier tracking
\cite{t_humphreys08_ptl,t_humphreys08_sis}, PpRx was designed from the
beginning for robust carrier recovery.  Likewise, from its inception PpEngine
was targeted for the harsh urban environment.  Over the past few years,
development of PpRx, PpEngine, and the LDRN has proceeded as a parallel
evolution, with each subsystem benefiting from improvements in the others.

The overriding design philosophy of this development has been to adapt,
rebuild, and reconfigure all three subsystems, separately and in parallel,
with the goal of minimizing $d_{95}$ while maximizing $P_V$, or, relatedly,
maximizing $P_S$ subject to $P_F \leq \bar{\bar{P}}_F$.  This approach
benefits greatly from a purely software-based approach to GNSS signal
processing (as opposed to processing that exploits dedicated silicon or
FPGAs), for two reasons.  First, a software-defined approach is almost
infinitely flexible: all processing downstream from the RF front end can be
reconsidered, rebuilt, and re-evaluated in a rapid iterative process using an
efficient and common high-level programming language.  Second,
software-defined receivers can exploit multiple cores to run faster
than real time on recorded IF samples\cite{t_humphreys09_sdgr}.  The
PpRx-PpEngine pipeline runs at 10x real time on a 6-core Intel Xeon 2.27 GHz
processor, enabling rapid iteration cycles for quickly probing the
optimization landscape.

\subsection{Carrier and code tracking}
\label{sec:carr-code-track}
\begin{figure}[!t]
\centering
\includegraphics[width = 8.5cm]{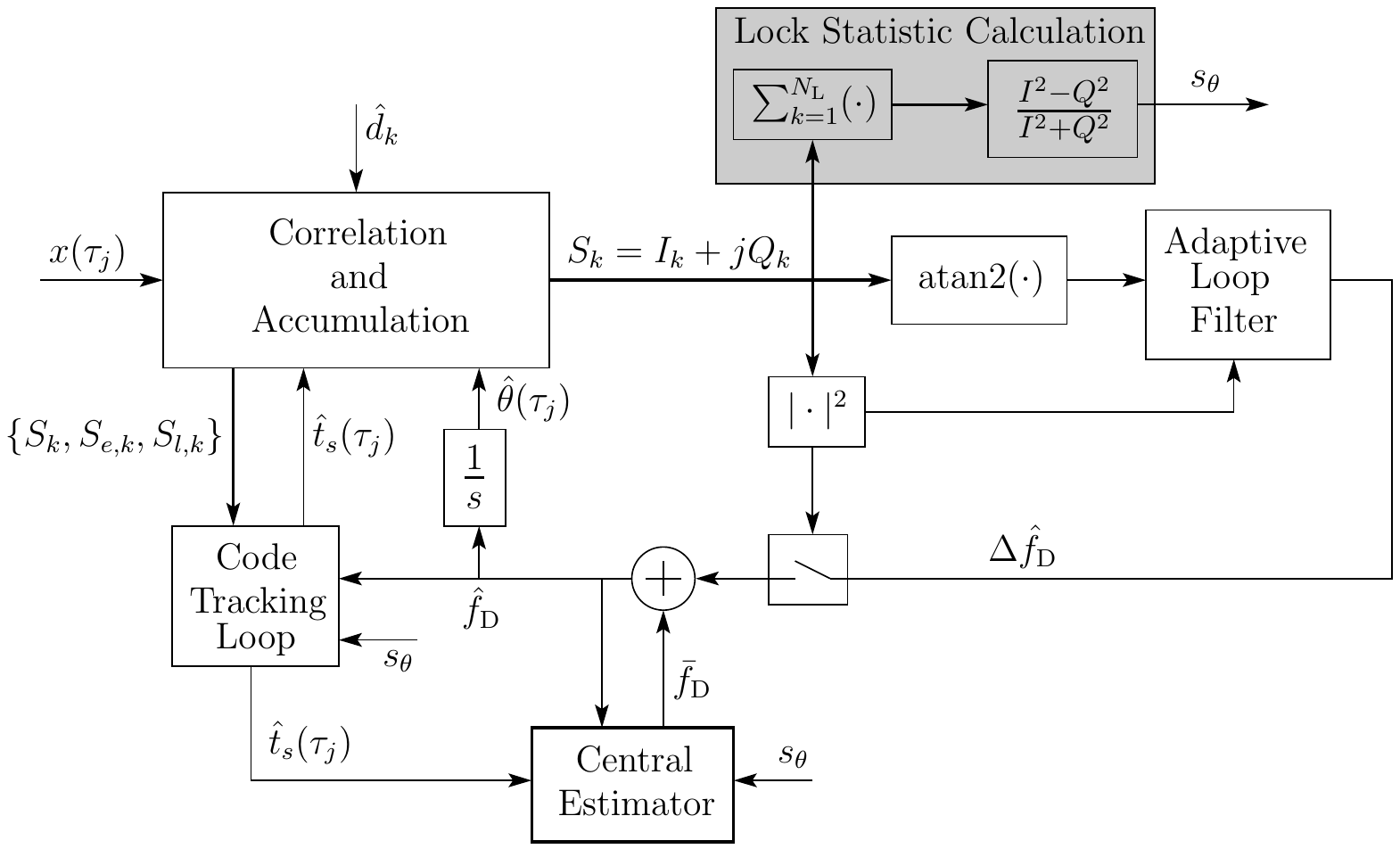}
\caption{PpRx's carrier and code tracking architecture.}
\label{fig:tracking}
\end{figure}

GNSS carrier and code tracking in an urban environment must be opportunistic,
taking advantage of short clear glimpses to overhead satellites as they
present themselves.  PpRx's code and carrier tracking architecture,
illustrated in Fig. \ref{fig:tracking} has been designed for immediate (within
approximately 100 ms) recovery of full-cycle phase lock after a blockage, and,
importantly, for prompt lock indication.  The following subsections describe
the essential elements of PpRx's tracking strategy, calling out parameters
whose values significantly affect urban RTK performance.

\subsubsection{Correlation and accumulation}
Correlation and accumulation is performed on a sequence of noisy IF samples
$x(\tau_j), ~ j = 0, 1, ...$, where $\tau_j$ denotes the time of the $j$th
sample according to the receiver's clock.  Within the correlation and
accumulation block, a complex local replica signal is formed with code and
carrier phase estimates $\hat{t}_s(\tau_j)$ and $\hat{\theta}(\tau_j)$
provided by the code and carrier phase tracking loops.  The outputs of the
correlation and accumulation block are prompt, early, and late complex
correlation products $S_k$, $S_{e,k}$, and $S_{l,k}$ of the form
$S_k = I_k + jQ_k$, where $I_k$ and $Q_k$ are the in-phase and quadrature
accumulations. (The green and blue traces in Fig. \ref{fig:iq_gps_prn4P}
correspond to $I_k$ and $Q_k$, respectively.) The accumulation interval,
$T_a$, is an important configuration parameter for urban RTK.

\subsubsection{Navigation data bit wipeoff}
\label{sec:lnav-data-bit}
The GPS L1 C/A and SBAS L1 signals have no dedicated pilot component.  The
phase ambiguity introduced by their full suppressed-carrier binary modulation
makes it challenging to recover an accurate carrier phase measurement in an
urban environment.  However, the reference network can provide low-latency
estimates $\hat{d}_k$ with which the incoming modulation can be ``wiped off,''
allowing full-cycle carrier recovery.

GPS L1 C/A data bit wipeoff has been employed for years to improve
weak-signal acquisition in smartphones \cite{van2009gps}, but, so far as the
authors are aware, it has not been previously applied in the context of CDGNSS
positioning.  SBAS L1 data wipeoff, a novel technique introduced in this
paper, is even more valuable on a per-signal basis than GPS L1 C/A data
wipeoff, since the short 2-ms binary SBAS symbol period otherwise renders SBAS
signals of little use for urban precise positioning. 

\subsubsection{Lock statistic calculation}
Also key to robust urban RTK is the ability to exclude corrupt or otherwise
inaccurate carrier phase measurements.  However, due to poor signal
availability, an urban RTK engine cannot afford to be overly conservative: it
must minimize the number of adequate-quality measurements that get falsely
labeled as corrupt.  An important indicator for this wheat-from-tares
separation is the lock statistic $s_\theta$.  Let $I$ and $Q$ be
coherent sums of $I_k$ and $Q_k$ over $N_{\rm L}$ accumulation intervals.
Then $s_\theta$ is calculated as \cite{a_vandierendonck96_gr}
\[ s_\theta = \frac{I^2 - Q^2}{I^2 + Q^2} \] The goal of the carrier tracking
loop is to adjust its phase estimate $\hat{\theta}_k$ to shift signal power
from $Q_k$ to $I_k$.  Thus, for a loop in lock, $I^2 >> Q^2$ and $s_\theta$ is
near unity.

A new lock statistic is produced every $N_{\rm L}$ accumulations.
$N_{\rm L}$, must be chosen large enough to suppress thermal noise in $I_k$
and $Q_k$, but small enough to provide a prompt indicator of phase lock to all
dependent processing.  PpEngine relies crucially on $s_\theta$ to screen out
bad measurements.  Note from Fig. \ref{fig:tracking} that $s_\theta$ is also
fed to the code tracking loop and to PpRx's central state estimator: each one
adapts its behavior to rely less on Doppler measurements when $s_\theta$ is
low.

\subsubsection{Carrier tracking}
\label{sec:carrier-tracking}
As illustrated in Fig. \ref{fig:tracking}, PpRx employs a vector signal
tracking architecture wherein a central estimator, implemented as a Kalman
filter with a nearly-constant-velocity dynamics model, receives observables
from all tracking channels and drives local replica generation for each
channel \cite{lashley2009vector}.  More particularly, PpRx employs a hybrid
strategy in which, for each channel, a local phase tracking loop is closed
around a modeled Doppler value $\bar{f}_{\rm D}$ provided by the central
estimator.  The local loop's residual Doppler frequency estimate
$\Delta \hat{f}_{\rm D}$ is added to $\bar{f}_{\rm D}$ to produce the full
estimate $\hat{f}_{\rm D}$ used in replica generation.

A four-quadrant arctangent phase discriminator, ${\rm atan2}(Q_k,I_k)$, which
is nearly optimal for decision-directed carrier recovery, and optimal for
data-free signals, or when data bit wipeoff is error-free, feeds a phase error
measurement at every accumulation interval to the carrier tracking loop
filter.  PpRx's carrier loop filter is designed according to the
controlled-root formulation of \cite{s_stephens95_pll}.  The filter adapts its
bandwidth $B_\theta$ at every accumulation interval according to the value of
$|S_k|$.  The adaptation schedule has a significant effect on RTK performance.

One might expect that adapting $B_\theta$ so to maintain a constant loop SNR
as $|S_k|$ varies would yield the best results.  This is effectively the
adaptation schedule that gets applied in Kalman-filter-based weak signal
tracking \cite{m_psiaki02_wgs}.  However, this reasonable approach was found
to yield reduced urban RTK performance.  More effective is a three-tiered
schedule that reduces $B_\theta$ when $|S_k|$ falls below a fairly low
threshold $\gamma_1$, and sets $B_\theta$ to zero if $|S_k|$ falls below
another threshold $\gamma_0 < \gamma_1$. Within this lowest tier,
$\Delta \hat{f}_{\rm D}$ is also driven to zero over a few accumulation
intervals, thereby breaking the local feedback loop.  In this open-loop mode,
the local replica's phase estimate is driven entirely by the model Doppler
$\bar{f}_{\rm D}$.  The lock statistic $s_\theta$ continues to be calculated.
If $s_\theta$ is sufficiently close to unity, the central estimator, the code
tracking loop, and the RTK engine continue to treat $\hat{\theta}(\tau_j)$ as
a valid measurement.  But this is a rare occurrence; $s_\theta$ is typically
far from unity in open-loop mode.

Such open-loop tracking has been found to be useful for preventing frequency
unlock during intervals when signals are entirely blocked, e.g., by buildings
or bridges, and for enabling fast re-acquisition of carrier lock immediately
following the blockage.

\subsubsection{Code tracking}
PpRx's code tracking loop, which is aided by the Doppler estimate
$\hat{f}_{\rm D}$, is implemented as a $1$st-order loop that toggles between a
non-coherent (dot product) discriminator and a coherent discriminator.  The
coherent discriminator is applied when the channel is phase locked and no
recent phase trauma (indicated by $s_\theta$) has been detected; otherwise,
the non-coherent discriminator is applied.  A flag attached to each code phase
measurement $\hat{t}_s(\tau_j)$ indicates to downstream processes whether it
was produced under coherent or non-coherent tracking.

As with carrier tracking, the code tracking loop filter's bandwidth,
$B_{t_s}$, is adaptive.  But rather than responding to $|S_k|$ as the carrier
loop's bandwidth does, $B_{t_s}$ takes on a different value for each of four
code tracking modes: (1) pre-phase lock, (2) first post-lock transient, (3)
second post-lock transient, and (4) steady-state.  These modes are designed to
ensure rapid convergence of the code phase estimate $\hat{t}_s(\tau_j)$ after
initial signal acquisition, or in the aftermath of phase unlock.

\subsection{Precise positioning}
PpRx and the LDRN send carrier and code phase observables, together with
signal quality indicators $s_\theta$ and $C/N_0$, and various other meta-data,
to PpEngine for processing.  PpEngine is capable of processing observables
from both rover antennas simultaneously, exploiting the known distance between
these.  But for the results presented in this paper, PpEngine was invoked only
in its simplest single-antenna mode, producing a precise 3-dimensional
baseline between the primary rover antenna and a selected reference station
antenna in the LDRN.  This simple single-baseline RTK mode was chosen so that
the precise positioning system's performance could be evaluated in a familiar
configuration and easily compared with other single-baseline RTK evaluations
such as \cite{li2018high}.

\subsubsection{Treatment of real- and integer-valued states}
The current embodiment of PpEngine adopts a straightforward approach to RTK.
It first forms code and carrier measurement double differences (DDs) from the
rover and reference data, then sends these to a mixed real/integer extended
Kalman filter for processing.  The filter is implemented as a square-root
information filter, as in \cite{psiakiFiltering2010}, but limits growth of the
number of integer states by either (1) marginalizing at each epoch over
float-valued integer ambiguity states modeled as Gaussian-distributed, or (2)
conditioning on the estimated integer values.  Thus, PpEngine's current
approach is to discard all integer states, by marginalization or by
conditioning, after each measurement epoch.  The marginalization option, which
yields the float solution, can be thought of as a special case of the
sub-optimal filter in \cite{psiakiFiltering2010} with a window length $i = 1$.
The conditioning option, which yields the fixed solution, is invoked only if
the integer estimates, found by integer least squares (ILS)
\cite{teunissen1995LAMBDA_journal}, are validated by an aperture test.

Conditioning the real-valued states on the lowest-cost integer estimates
yields a maximum \emph{a posteriori} 3D baseline estimate.  After each
measurement update, the real-valued states are propagated to the next
measurement epoch, whereupon a new set of integer estimates are formed and
conditioning or marginalization occurs yet again.  Importantly, if the integer
states are validated at the $l$th measurement epoch, it is the
integer-conditioned real-valued states that are propagated to the $(l+1)$th
measurement epoch.  Thus, although all integer states are discarded between
measurement updates, correct integer resolution is highly likely at the
$(l+1)$th epoch if integer ambiguities were correctly resolved at the $l$th
epoch because the real-valued states carry forward a decimeter-accurate
position estimate.

Carrying forward integer-conditioned real-valued states is perilous because
eventually an erroneous integer estimate passes the aperture test, whereupon
the integer-conditioned real-valued states are corrupted by conditioning on
the incorrect fix.  What is more, the associated square-root information
matrices indicate high confidence in the erroneous real-valued state, raising
the chances that the next integer estimates, which are constrained by the
prior real-valued states, will also be incorrectly fixed.  This cycle, which
can persist for several seconds, is eventually broken by an aperture test
failure prompted by signal loss, large measurement errors, or the persistent
lack of consistency between the incoming observables and the current state.

In view of this peril, the authors are developing a generalization of PpEngine
that can manage growth in the number of integer state elements using a variant
of the suboptimal approach of \cite{psiakiFiltering2010}.  Meanwhile,
PpEngine's single-epoch integer resolution has the virtue of being insensitive
to cycle slips that occur between measurement epochs, which are common in the
urban environment.

\subsubsection{Dynamics Model} Because this paper's focus is on RTK unaided by
any non-GNSS sensors, the mixed real- and integer-valued state estimator
within PpEngine was configured to ignore all available inertial measurements
and instead rely on a simple nearly-constant-velocity dynamics model for state
propagation between measurements.  The dynamics model assumes roughly
equivalent process noise variance in the along-track and cross-track
directions, but smaller variance (by a factor of 100) in the vertical
direction, in keeping with a land vehicle operating in a relatively flat urban
environment.

\subsubsection{Robust measurement update}
\label{sec:robust-meas-update}
Urban multipath and diffraction cause code and carrier observables to exhibit
large errors with a much higher probability than even a conservative Gaussian
model would predict.  Dealing with measurement error processes such as these,
which have thick-tailed distributions, requires robust estimation techniques;
that is, techniques with reduced sensitivity to measurement outliers.

\begin{figure}[!t]
\centering
\includegraphics[width = 7cm]{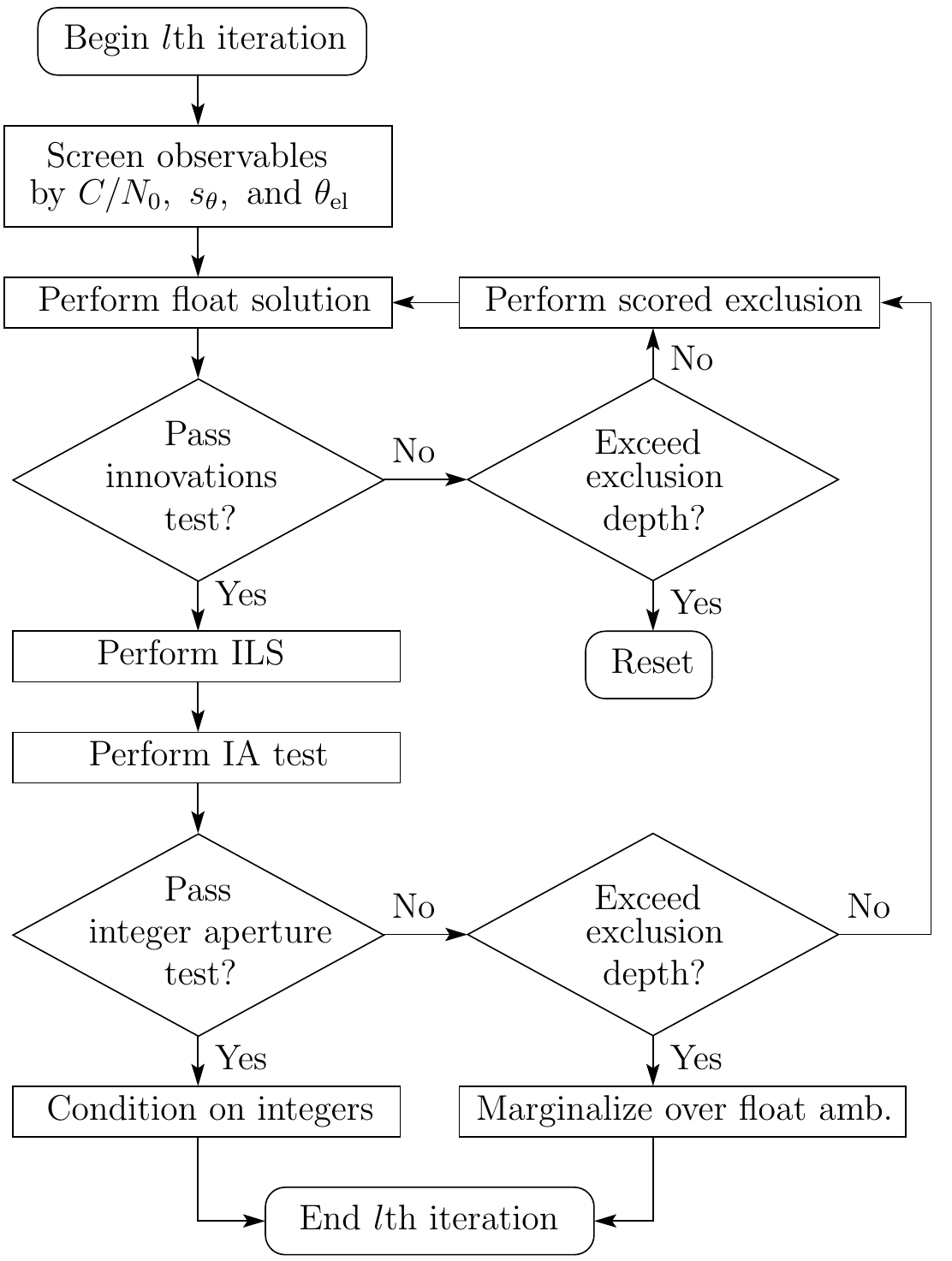}
\caption{Flow diagram for the PpEngine exclusion and fixing logic.}
\label{fig:rtkFlow}
\end{figure}

Outliers are especially problematic for integer fixing in RTK positioning.  By
action of the decorrelation adjustment preceding ILS, a single bad measurement
can contaminate multiple measurements in the decorrelated domain, rendering
resolution of the associated integers impossible.  Partial ambiguity
resolution, as in \cite{brack2015giafit,parkins2011increasing}, offers little
relief in such cases because contamination caused by outliers is not
necessarily limited to an identifiable subset of integers.  It is more
effective to exclude questionable measurements before the decorrelation
adjustment.

PpEngine implements a multi-level exclusion process, depicted in
Fig. \ref{fig:rtkFlow}, to mitigate the effects of measurement outliers.  At
each measurement epoch, measurements are first screened on the basis of three
quality indicators: carrier-to-noise ratio $C/N_0$, phase lock statistic
$s_\theta$, and elevation angle $\theta_{\rm el}$.  Signals whose values fall
below user-selected thresholds for these quantities are excluded from all DD
combinations.

A second level of exclusion occurs as part of the float solution.  A
$\chi^2$-type test is applied to all DD measurement innovations
\cite{y_barshalom01_tan}, with exclusion triggered if the normalized
innovations squared statistic exceeds a chosen threshold.  For the current
implementation of PpEngine, this test is only effective at excluding anomalous
DD code phase (pseudorange) measurements, since the float states are
discarded, and thus unconstrained, from epoch to epoch.  Note that innovations
testing benefits strongly from a correctly integer-constrained state because
the exclusion threshold can be made tighter.  However, with an
incorrectly-integer-constrained state, innovations testing may end up
excluding the very measurements necessary to correct the state.

If a set of innovations fails the innovations test, DD measurements (both code
and carrier for a particular DD combination) are excluded one at a time (with
replacement).  Exclusion is ordered such that the next DD combination removed
is the one with the next-lowest quality score that has not yet been removed.
A quality score is formed for each DD combination via a linear combination of
scores based on $C/N_0$, $s_\theta$, and $\theta_{\rm el}$.  If such
$N$-choose-$1$ elimination fails to create a subset of DD measurements that
passes the innovations test, exclusion can proceed to $N$-choose-$m$
elimination, with $m > 1$.  If a user-configurable exclusion depth is
exceeded, the estimator state is reset.

The third level of exclusion is based on the integer aperture test following
integer estimation via ILS.  This is the standard data-driven integer fixing
process whereby the integer-fixed solution is selected only on successful
validation by some type of aperture test; otherwise, the float solution is
accepted \cite{green2018iagiab}.  The aperture test is configured for a fixed
failure rate (under independent Gaussian errors) of $\bar{P}_F$.  If the
integer aperture test fails, $N$-choose-1 exclusion (with replacement) is
attempted, starting with the lowest-scoring DD combinations and working up
through higher-scoring combinations. $N$-choose-$m$ exclusion, with $m > 1$,
is currently not attempted at this layer of exclusion because testing a large
number of subsets is eventually ``doomed to succeed'' at passing the aperture
test, causing $P_{F}$ to significantly exceed $\bar{P}_F$ even under benign
conditions \cite{parkins2011increasing}.

If the aperture test is passed before the permissible exclusion depth is
exceeded, the solution is conditioned on the integers and the integer states
are dropped.  Otherwise, the integer state elements are marginalized out as
float values.  In either case, the state is propagated to the next measurement
epoch via the dynamics model and the process repeats.

\section{Experimental Setup}
The precise positioning system was evaluated experimentally using data
collected on Aug. 1, 2018 during approximately 2 hours of driving in and
around the dense urban center of Austin, TX.

\begin{figure}[!t]
\centering
\includegraphics[width = 8.5cm]{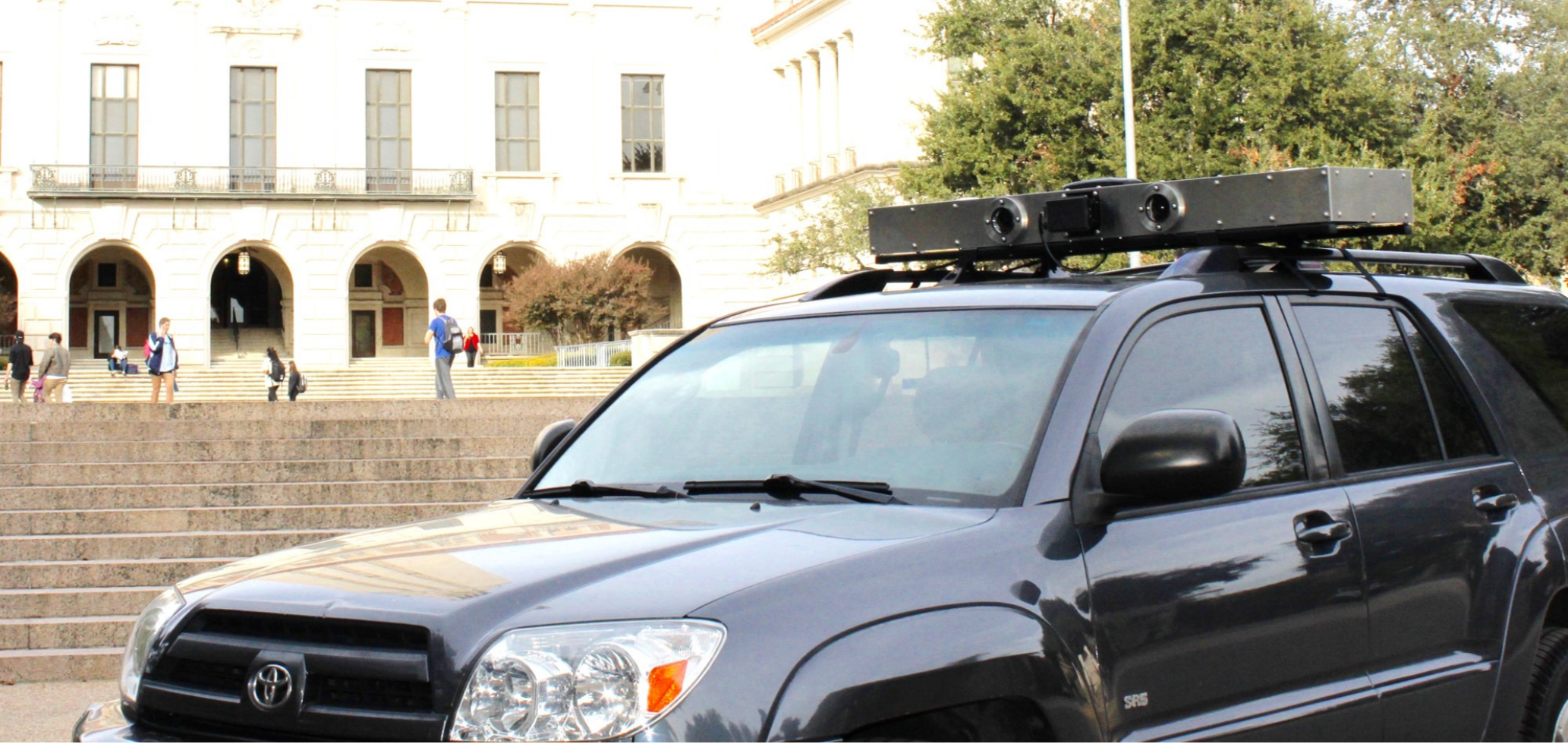}
\caption{The University of Texas Sensorium is a platform for automated and
  connected vehicle perception research.  It includes stereo visible light
  cameras, an industrial-grade IMU, an automotive radar unit, a dual-antenna,
  dual-frequency software-defined GNSS receiver, 4G cellular connectivity, and
  a powerful internal computer.}
\label{fig:sensorium}
\end{figure}

The rover GNSS receiver is one among several sensors housed in an integrated
perception platform called the Sensorium, pictured in
Fig. \ref{fig:sensorium}.  Although hardly visible in
Fig. \ref{fig:sensorium}, two Antcom G8 triple-frequency patch antennas are
flush-mounted in the cross-track direction on the Sensorium's upper plate,
separated by just over 1 meter.  The antennas' signals are routed to a unified
RF front end whose output IF samples are processed in real time (to within
less than 10 ms latency) by the Sensorium's onboard computer.  The samples are
also stored to disk for post-processing.

Data from both the driver- and passenger-side antennas were used to produce
the PpRx standard navigation solution, but only data from the driver-side
antenna were used in the urban RTK performance evaluation.  No other Sensorium
sensors were involved in the current paper's results.

\begin{figure}[!t]
\centering
\includegraphics[width = 8.5cm]{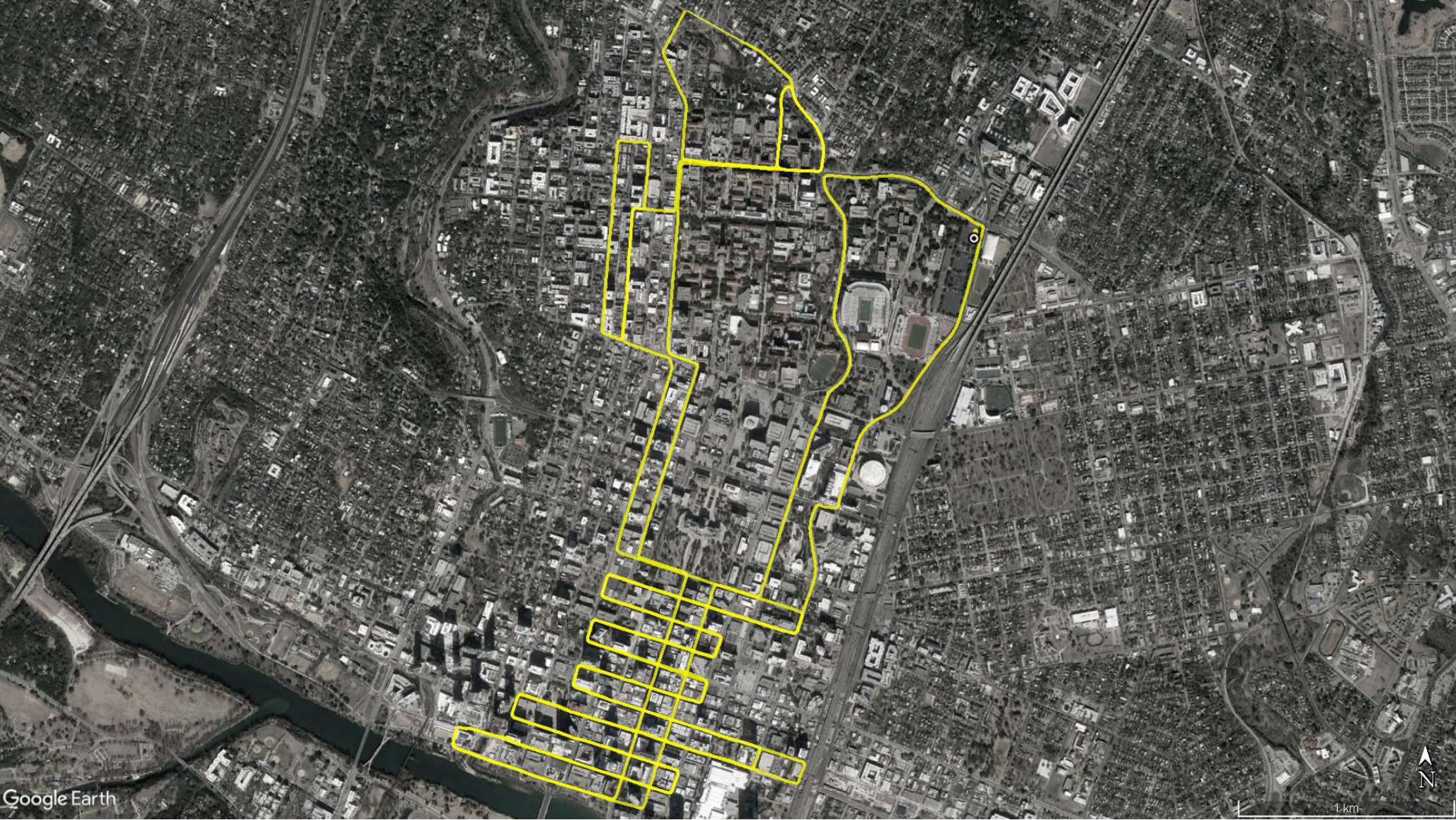}
\caption{Overview of the test route through the urban core of Austin, TX.}
\label{fig:routeOverview}
\end{figure}

The test route, depicted in Fig. \ref{fig:routeOverview}, runs the gamut of
light-to-dense urban conditions, from open-sky to narrow streets with
overhanging trees to the high-rise urban city center. A time history of route
coordinates, in the form of a Google Earth KML file, is packaged with the
other campaign data so that readers can explore the route.  The route begins
with a 10-minute, and ends with a 4-minute stationary interval in open sky
conditions to allow confident bookending for the ground truth system.  The
number $N_{\rm DD}$ of double-difference signals available to PpEngine at each
epoch over the 2-hour test interval ranged from 1 to 18, with an average of
12.5.

\section{Ground Truth Trajectory}
A trustworthy ground truth trajectory against which to compare the reported
trajectory of the system under test is indispensable for urban positioning
evaluation.  The ground truth generation process of
\cite{humphreys2018urbanStandAloneRTKplans2018} was unsatisfactory for two
reasons.  First, it lacked independence, as it drew in part on the same
underlying precise solutions that were to be evaluated.  Second, it was not
possible to create a complete ground truth trajectory even for the moderate
urban test route of \cite{humphreys2018urbanStandAloneRTKplans2018}.  Gaps in
the ground truth prevented an accurate determination of $P_F$.

The present work adopts the more traditional approach of taking the
forward-backward smoothed trajectory generated in after-the-fact processing by
a coupled RTK-inertial system with a tactical-grade IMU as the ground truth
\cite{ong2009assessment,li2018high}.  In particular, an iXblue ATLANS-C mobile
mapping INS/GNSS system, which incorporates a professional-grade Septentrio
AsteRx3 RTK receiver, was used to generate the ground truth
\cite{ixblueAtlansCDatasheet}.  The ATLANS-C was rigidly mounted to the
Sensorium and attached to the same antenna from which PpEngine drew
observables.  A cm-accurate lever arm estimate from the inertial sensor to the
GNSS antenna was determined.  Self-reported 3D accuracy of the ATLANS-C's
smoothed estimate varied between 2 and 20 cm (1-sigma) along the test route.
Along the light-to-moderate urban portions of the test route, the ATLANS-C and
PpEngine 3D estimates agreed to better than 5 cm (95\%).

\section{Baseline System Performance}
The baseline urban RTK system is the PpRx-PpEngine pipeline configured to
maximize $P_S$ while respecting $P_F \leq \bar{\bar{P}}_F$ for some chosen
empirical incorrect fixing probability bound $\bar{\bar{P}}_F$.  This section
discusses the baseline system's configuration and performance.  The following
section will compare the baseline system against several
alternative configurations of the PpRx-PpEngine pipeline.

\subsection{Configuration}
PpRx's carrier and code tracking loops were configured as
detailed in \cite{humphreys2018urbanStandAloneRTKplans2018}.  PpRx was
configured to track the following signal types: GPS L1 C/A, GPS L2C (combined
M + L tracking), Galileo E1 BOC(1,1) (combined B + C tracking), and SBAS
(WAAS) on L1.  It was configured to output observables at 5 Hz.

PpEngine was configured as follows.  The master LDRN reference station,
located within 4 km of all points on the test route, was taken as the
reference receiver, producing reference observables at 5 Hz. The master
station's antenna is a Trimble Zephyr II geodetic antenna.  A single-baseline
RTK solution with a near-zero age of data was performed between the rover's
primary antenna and the reference station at a 5-Hz cadence.  The following
thresholds were applied in the first-level screening processing within
PpEngine: $C/N_0 \geq 37.5$ dB-Hz, $s_\theta \geq 0.5$, and
$\theta_{\rm el} \geq 15$ deg. Signals whose values fell below any one of
these thresholds were excluded from all DD combinations.  Elevation-dependent
weighting was applied in the float solution.  The threshold above which float
innovation statistics failed the normalized innovation squared test was chosen
to be 2.  Scored $N$-choose-1 exclusion was applied for both failed float
innovations tests and failed aperture tests.  A depth of 8 signals was allowed
for the $N$-choose-1 exclusion, after which the estimator was either reset or
integers marginalized, according to the flow diagram in
Fig. \ref{fig:rtkFlow}.  The difference test of \cite{wang2015iafit}, which
was found to work as well in urban environments, was chosen as the integer
aperture test.  The test was configured for a fixed failure rate of
$\bar{P}_F = 0.001$. The undifferenced pseudorange and phase measurement error
were taken to be $\sigma_\rho = 0.9$ m and $\sigma_\phi = 4$ mm,
respectively. The nearly-constant-velocity dynamics model was configured for a
0.4 m/s deviation in horizontal velocity, and a 0.06 m/s deviation in vertical
velocity over a 1-second interval.

A calibration was carried out of the Sensorium antennas' phase center
variation with elevation angle relative to the reference antenna.  The
calibration procedure is similar to the one presented in \cite{mader1999gps}
except that it works with double instead of single differences.  The
calibration succeeded in reducing the standard deviation of L1 and L2
undifferenced carrier phase residuals by 11\% and 15\%, respectively, in
open-sky conditions.

\subsection{Performance}
\begin{figure}[!t]
\centering
\includegraphics[width = 8.5cm]{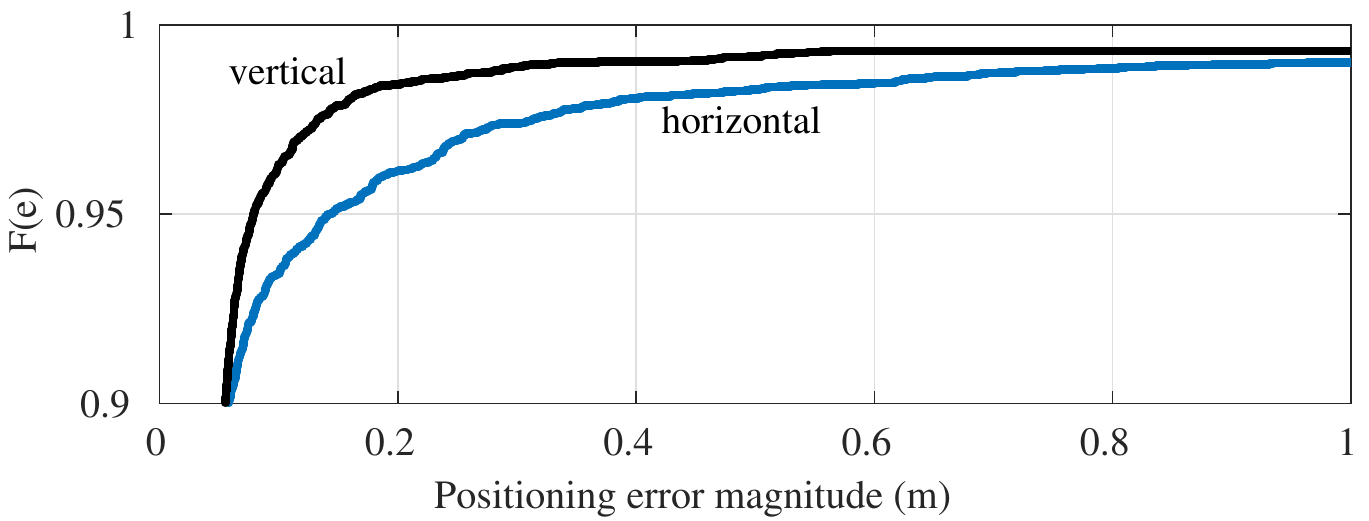}
\caption{Cumulative distribution function for horizontal and vertical fixed
  position error magnitudes with respect to the ground truth for the baseline
  system.}
\label{fig:errorCdfPlot_s1}
\end{figure}

Fig. \ref{fig:errorCdfPlot_s1} shows the cumulative distribution function
(CDF) of the horizontal and vertical positioning errors for fixed
(aperture-test-validated) PpEngine solutions.  Positioning performance appears
excellent, with 95\% of horizontal and vertical errors below 14 and 8 cm,
respectively.  That the vertical errors are smaller than the horizontal errors
is explained by the vehicle motion's greater vertical predictability.

\begin{figure}[!t]
\centering
\includegraphics[width = 8.5cm]{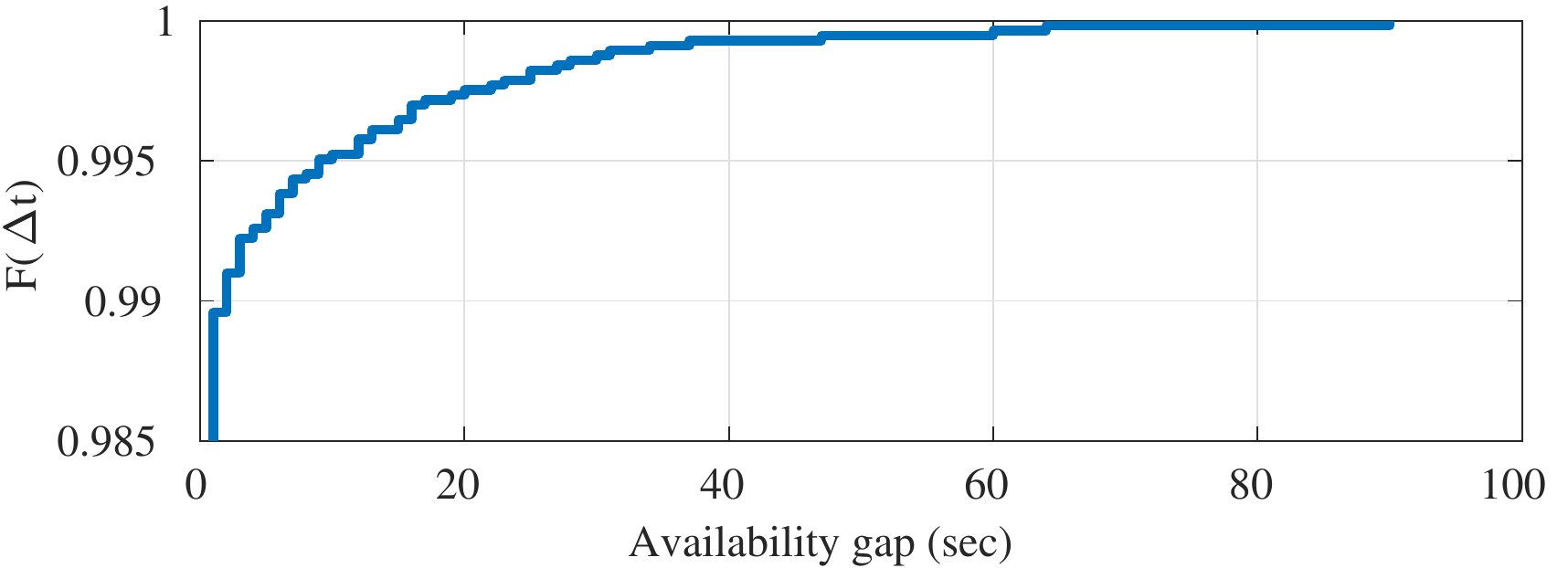}
\caption{Cumulative distribution function for baseline system availability
  gaps.}
\label{fig:availabilityGapCdfPlot_s1}
\end{figure}

Fig. \ref{fig:availabilityGapCdfPlot_s1} shows the CDF of availability gaps in
the baseline system's fixed solution.  These are intervals during which only a
less-accurate float solution is available.  Although the longest gap was 90
seconds, over 99\% of gaps are shorter than 2 seconds, a span that could be
easily bridged by a MEMS-quality inertial sensor with errors smaller than a
few cm \cite{zhangComparisonWithTactical2006}.

The baseline system's fixed solution availability, $P_V$, was 87.2\%.  Fixed
solutions were considered correctly resolved if their 3D positions were within
30 cm of the ground truth.  This led to $P_S = 84.8\%$, and $P_F = 2.4\%$.
Note that $P_F$ is a factor of 24 larger than $\bar{P}_F = 0.1\%$ but may be
tolerable for a larger system that combines stand-alone RTK with inertial and
electro-optical sensing, as the Sensorium of Fig. \ref{fig:sensorium} is
intended to do.

\section{Performance Degradation Analysis}
This section presents a performance degradation analysis in which features of
the baseline system are removed or altered one at a time to assess their
relative contribution to baseline system performance.  Table
\ref{tab:summaryTable}, where $P_V$, $P_S$, and $P_F$ are as defined
previously, summarizes the results of the analysis.  Starting with Scenario 2,
the following discussion treats each scenario in turn.

\begin{table*}[!t]
  \centering
  \caption{Summary of precise positioning results}
  \label{tab:summaryTable}
  \begin{tabular}[c]{clccc}
    \toprule
    Scenario &  Description & $P_{V}$: Validated Epochs (\%)
    & $P_S$: Success (\%) & $P_F$: Failure (\%)  \\ \midrule
    1 & Baseline system & 87.2 & 84.8 & 2.4 \\ 
    2 & LNAV \& SBAS data bit prediction disabled & 79.4 & 54.4 & 25.0 \\  
    3 & Scalar tracking with adaptive $B_\theta$ & 87.1   & 84.1 & 3.0 \\  
    4 & Scalar tracking with fixed $B_\theta$ & 86.9 & 80.7 & 6.2 \\ 
    5 & GPS L2CL tracking & 83.2 & 80.8 & 2.3 \\ 
    6 & Age of data = 200 ms & 87.4 & 85.0 & 2.4 \\ 
    7 & Age of data = 400 ms & 87.3 & 83.8 & 3.5 \\  
    8 & Age of data = 600 ms & 87.2 & 82.1  & 5.1 \\  
    9 & Age of data = 1000 ms & 87.0 & 82.0  & 5.0 \\ 
    10 & 15 km baseline & 86.6 & 78.9  & 7.7  \\  
    11 & Sans SBAS & 78.6 & 73.1  & 5.5 \\  
    12 & Sans GPS L2C (L+M) & 83.6 & 82.5  & 1.1 \\  
    13 & Sans Galileo E1 (B+C) & 77.4 & 75.9  & 1.4 \\  
    14 & No scored exclusion & 78.8 & 75.6 & 3.2 \\ 
    15 & No antenna calibration & 87.0 & 82.8  & 4.2 \\  
    \bottomrule
  \end{tabular}
\end{table*}

\paragraph*{Data bit prediction disabled}
Eliminating the baseline's system's LNAV and SBAS data bit prediction
capability, described in Section \ref{sec:lnav-data-bit}, has a devastating
effect on performance.  The availability of validated epochs drops by 8
percent points and $P_F$ rises tremendously, from 2.4\% to 25\%. Clearly, data
bit prediction is a key capability for urban RTK.

\paragraph*{Scalar tracking with adaptive $B_\theta$}
Eliminating vector tracking, as described in Section
\ref{sec:carrier-tracking}, in favor of scalar tracking, but retaining carrier
tracking loop bandwidth adaptation, has no significant effect on $P_V$ but
$P_F$ increases slightly, from 2.4\% to 3\%.  Thus, vector tracking appears
helpful, but not critically so.

\paragraph*{Scalar tracking with fixed $B_\theta$}
Eliminating both vector tracking and carrier tracking loop bandwidth
adaptation has little effect on availability, but $P_F$ rises to 6.2\%,
indicating that loop bandwidth adaptation is useful in preventing fixing
errors.

\paragraph*{GPS L2CL tracking}
For GPS L2C tracking, PpRx jointly tracks the pilot (CL) and data-bearing
medium-length (CM) codes, wiping off the INAV data symbols modulating the CM
code with symbol value estimates based not on prediction, as with LNAV, but
merely on observation.  The rationale for this strategy is that the CL pilot
renders prediction less necessary than for the GPS L1 C/A signal, which does
not enjoy a pilot.  Eliminating joint L2C L+M tracking in favor of pure L2CL
tracking might be thought a more reliable strategy given that no symbol
wipeoff mistakes are ever made when tracking only the pilot.  However, Table
\ref{tab:summaryTable} indicates that this leads to a drop in availability
with hardly any improvement in the error rate.  Thus, it appears that joint CL
and CM tracking is preferred.

\paragraph*{Age of data}
Scenarios 6-9 explore the effect of increased age of reference data, from the
baseline age (near zero latency relative to the rover stream) to 1 second.
Little reduction occurs in $P_V$, but there appears a somewhat steady increase
in $P_F$ after 200 ms.

\begin{figure}[!t]
\centering
\includegraphics[width = 8.5cm]{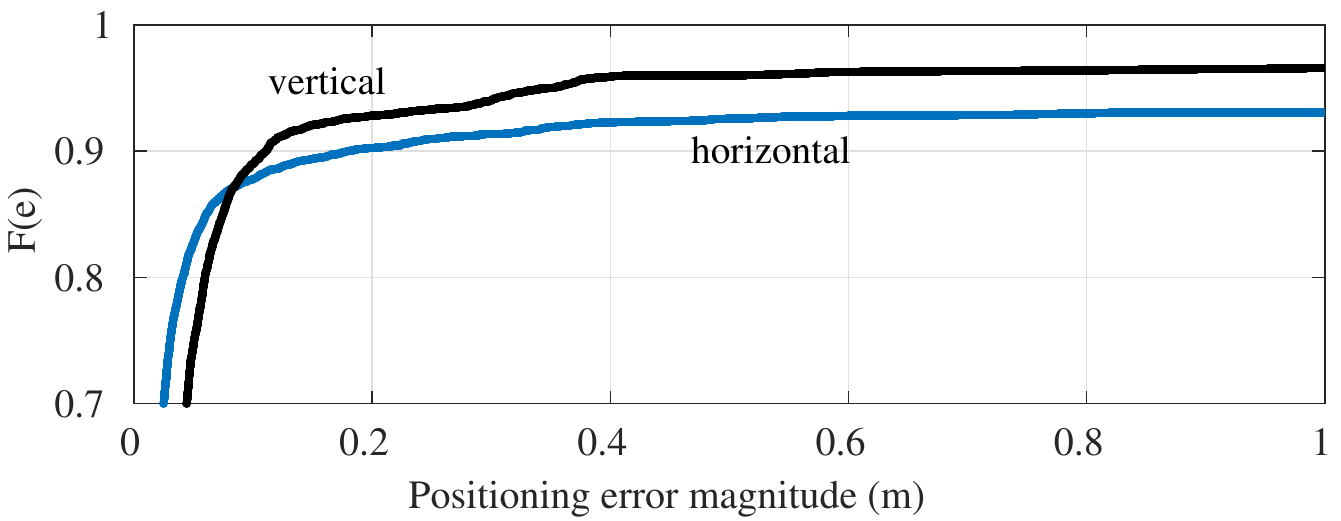}
\caption{Cumulative distribution function for horizontal and vertical
  fixed position error magnitudes with respect to the ground truth for Scenario
  10.}
\label{fig:errorCdfPlot_s10}
\end{figure}

\paragraph*{15-km baseline}
The baseline system's distance to the reference receiver, referred to as the
reference-rover baseline, is no greater than 4 km.  For Scenario 10, the LDRN
alternate master station, which sits 15 km from the furthest portion of the
test route, was instead taken as reference.  The alternate master station has
a Trimble Zephyr II antenna identical to the master station's.  A 15-km
baseline might still be considered within the short-baseline regime for
standard RTK \cite{odolinski2015combined}.  Nonetheless, a slight decrease in
$P_V$ and a significant increase in $P_F$ is observed, consistent with the
argument in \cite{murrian2016lowCostForAutomatedGpsWorld} that a dense
reference network is helpful in urban settings with reduced signal
availability.  Fig. \ref{fig:errorCdfPlot_s10} also shows a significant
degradation in overall fixed position accuracy.

\paragraph*{Value of additional signals}
Scenarios 11-13 explore the degradation that occurs when all signals of a
particular type are eliminated from the RTK solution.  Curiously, dropping L2C
and Galileo from consideration significantly reduced $P_F$, likely due to the
misfortune that both groups were composed primarily of low-elevation
satellites (below 30 degrees) during the test run.  By contrast, one notes a
significant increase in $P_F$ as the 4 available SBAS satellites are removed
from consideration.  Clearly, data-wiped SBAS signals offer significant
strength to the solution.

\paragraph*{Scored Exclusion}
Scenario 14 removes the scored exclusion strategy described in Section
\ref{sec:robust-meas-update} by setting the exclusion depth to 0.  This caused
a noticeable reduction in $P_V$ and a slight increase in $P_F$.

\paragraph*{Antenna calibration}
Scenario 15 indicates that lack of rover antenna calibration has no
discernible effect on $P_V$, but increases $P_F$ significantly.  The effect
would no doubt be larger for lower-quality rover antennas.

\section{Performance Comparison}
To further assess its performance, the PpRx-PpEngine system was compared
against three alternative systems: (1) A high-end commercial RTK system: a
Septentrio AsteRx3 receiver attached to the same GNSS antenna as the
PpRx-PpEngine system, with RTK solutions produced by Septentrio's RTK engine,
taking reference data from the CORS station TXAU, which sits less than 5.4 km
from the furthest portion of the test route; (2) a state-of-the-art PPP
solution from the CSRS-PPP service \cite{csrs2019ppp} (version v2.26.0 from
March 2019) based on observables produced by the same receiver as in (1)
processed in a ``batch kinematic'' mode; and (3) the so-called enhanced code
phase positioning (ECPP) solution \cite{narula2018accurate} from PpRx, which
draws on precise orbit and clock models from the IGS \cite{igs2018products}, a
WAAS ionospheric model, and RF signals from both passenger- and driver-side
Sensorium antennas.

Note that the receiver in (1) and (2) is the very receiver used to generate
the ground truth trajectory, but for this comparison its data were processed
without aiding from the ATLANS-C IMU. Also, to ensure a fair comparison,
PpEngine exclusion and validation tests were tightened (at the expense of
solution availability) until its probability of incorrect fix, $P_F$, was less
than that of the commercial RTK system.

Table \ref{tab:comparisonResultsTable} shows the comparison results in terms
of solution availability (equivalent to $P_V$ for the PpRx-PpEngine and
Septentrio RTK systems), $P_F$ (which applies only to the PpRx-PpEngine and
Septentrio RTK systems, as the other two do not attempt integer fixing), and
$d_{95h}$, the horizontal 95th percentile positioning error.  Significantly,
PpRx-PpEngine enjoys a 16\% availability advantage over the commercial RTK
system, and neither the CSRS-PPP nor the ECPP solutions are close to sub-30-cm
accuracy.

\begin{table}[!t]
  \centering
  \caption{Comparison results}
  \label{tab:comparisonResultsTable}
  \begin{tabular}[c]{lccc}
    \toprule
    System & Availability (\%) & $P_F$: Failure (\%) & $d_{95h}$ (cm)  \\ \midrule
    PpRx-PpEngine & 77 & 0.7 & 7.5 \\
    Commercial RTK & 61 & 0.8 & 8.6 \\
    CSRS-PPP & 74 & N/A & 154 \\                          
    ECPP & 100 & N/A & 275 \\
    \bottomrule
  \end{tabular}
\end{table}

\section{Conclusions}
A real-time kinematic (RTK) positioning system tailored for urban vehicular
positioning has been described and evaluated.  To facilitate performance
comparison against similar systems, the system was tested without any benefit
of aiding by inertial or electro-optical sensors.  Over nearly 2 hours of
urban testing, including multiple passes through Austin's dense urban center,
the system achieved an 85\% probability of correct integer fix for a 2.4\%
probability of incorrect fix, resulting in 3D positioning errors smaller than
17 cm (95\%).  A performance sensitivity analysis revealed that navigation
data bit prediction on fully-modulated GNSS signals is key to high-performance
urban RTK positioning, and that a dense reference network, carrier tracking
bandwidth adaptation, and rover antenna calibration each offer a significant
integrity benefit. A comparison with existing unaided systems for urban GNSS
processing indicates that the proposed system has a significant advantage in
availability and/or accuracy.

\section*{Acknowledgments}
The authors wish to thank iXblue engineers Jean-Baptiste Lacambre and Tim
Barford for providing the ground truth trajectory based on the iXblue ATLANS-C
system.  This work has been supported by the National Science Foundation under
Grant No. 1454474 (CAREER), by the Data-supported Transportation Operations
and Planning Center (DSTOP), a Tier 1 USDOT University Transportation Center,
and by the University of Texas Situation-Aware Vehicular Engineering Systems
(SAVES) Center (http://utsaves.org/), an initiative of the Wireless Networking
and Communications Group.




%
%
%

\bibliographystyle{IEEEtran}
\bibliography{pangea}

%

\newpage

\begin{IEEEbiography}[{\includegraphics[width=1in,height=1.3in,clip,keepaspectratio]{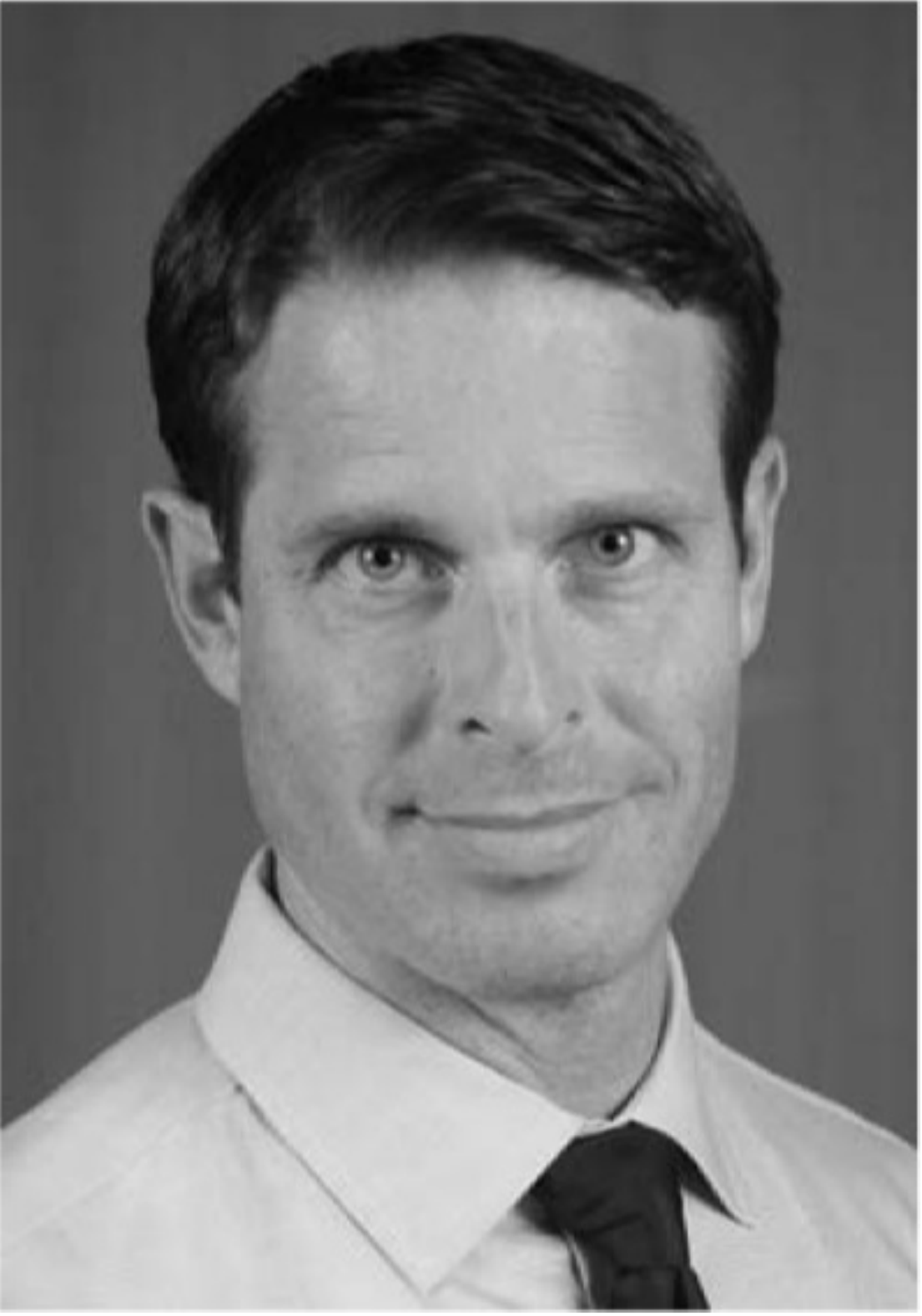}}]{Todd E. Humphreys}
  received the B.S. and M.S. degrees in electrical and computer engineering
  from Utah State University, Logan, UT, USA, in 2000 and 2003, respectively,
  and the Ph.D. degree in aerospace engineering from Cornell University,
  Ithaca, NY, USA, in 2008.

  He is an Associate Professor with the Department of Aerospace Engineering
  and Engineering Mechanics, The University of Texas (UT) at Austin, Austin,
  TX, USA, and Director of the UT Radionavigation Laboratory. He specializes
  in the application of optimal detection and estimation techniques to
  problems in satellite navigation, autonomous systems, and signal
  processing. His recent focus has been on secure perception for autonomous
  systems, including navigation, timing, and collision avoidance, and on
  centimeter-accurate location for the mass market.

  Dr. Humphreys received the University of Texas Regents' Outstanding Teaching
  Award in 2012, the National Science Foundation CAREER Award in 2015, and the
  Institute of Navigation Thurlow Award in 2015.
\end{IEEEbiography}
\begin{IEEEbiography}[{\includegraphics[width=1in,height=1.3in,clip,keepaspectratio]{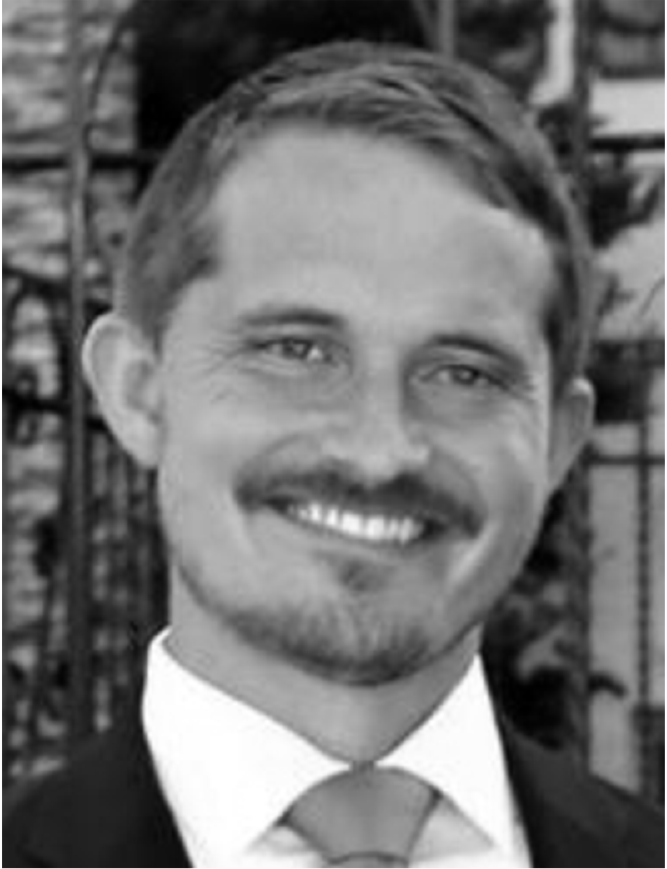}}]{Matthew J. Murrian}
served as a U.S. Navy submarine officer before joining the Radionavigation
Laboratory at the University of Texas at Austin, where he obtained a Masters in
aerospace engineering in 2018. He is now a systems engineer at CTSi working on
advanced navigation.
\end{IEEEbiography}
\begin{IEEEbiography}[{\includegraphics[width=1in,height=1.3in,clip,keepaspectratio]{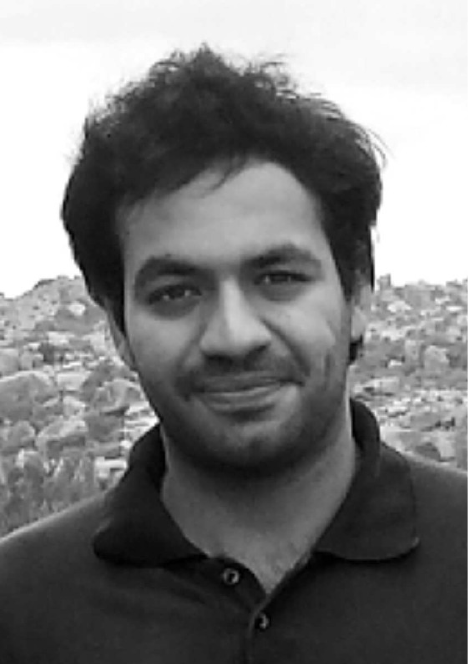}}]{Lakshay Narula}
  received the B.Tech. degree in electronics engineering from IIT-BHU, India,
  in 2014, and the M.S. degree in electrical and computer engineering from The
  University of Texas at Austin, Austin, TX, USA, in 2016.

  He is currently a Ph.D. student with the Department of Electrical and
  Computer Engineering at The University of Texas at Austin, and a Graduate
  Research Assistant at the UT Radionavigation Lab. His research interests
  include GNSS signal processing, secure perception in autonomous systems, and
  detection and estimation.

  Lakshay has previously been a visiting student at the PLAN Group at
  University of Calgary, Calgary, AB, Canada, and a systems engineer at Accord
  Software \& Systems, Bangalore, India. He was a recipient of the 2017
  Qualcomm Innovation Fellowship.
\end{IEEEbiography}



\vfill


\end{document}